\documentclass[mathematics,article,accept,moreauthors,pdftex]{Definitions/mdpi}

\usepackage[normalem]{ulem} 
\usepackage{amsfonts} 
\usepackage{wasysym} 
\usepackage{subfigure} 
\usepackage{mathcomp} 
\usepackage{CJKutf8} 
\usepackage{pifont} 
\usepackage{bm} 
\graphicspath{{./Definitions/}} 

\usepackage{cleveref}
\crefname{figure}{Figure}{Figures}
\crefname{specialtable}{Table}{Tables}
\crefrangelabelformat{figure}{#3#1#4--#5#2#6}
\crefrangelabelformat{specialtable}{#3#1#4--#5#2#6}
\crefname{section}{Section}{Sections}
\crefname{paragraph}{Section}{Sections}
\crefrangelabelformat{section}{#3#1#4--#5#2#6}
\crefname{appendix}{Appendix}{Appendices}
\crefrangelabelformat{appendix}{#3#1#4--#5#2#6}
\crefname{scheme}{Scheme}{Schemes}
\crefrangelabelformat{scheme}{#3#1#4--#5\crefstripprefix{#1}{#2}#6}
\crefname{chart}{Chart}{Charts}
\crefrangelabelformat{chart}{#3#1#4--#5\crefstripprefix{#1}{#2}#6}
\newcommand{\fig}[1]{Figure~\ref{#1}}
\newcommand{\sect}[1]{Section~\ref{#1}}

\makeatletter
\def\T@n@@nc@d@ngM@cr@M@d{}
\def\LY@n@@nc@d@ngM@cr@M@d{}
\makeatother

\let\orignewcommand\newcommand  
\let\newcommand\providecommand  
\usepackage{verse}
\let\newcommand\orignewcommand  

\newsavebox\foobox

\renewcommand{\dagger}{\mathchar"2279}
\renewcommand{\ddagger}{\mathchar"227A}

\newcommand{\mmathit}[1]{
  \ifthenelse{\equal{#1}{\ln}}{\mathit{ln}}{
    \ifthenelse{\equal{#1}{\max}}{\mathit{max}}{\mathit{#1}}
  }
}
\makeatother
\robustify{\footnote} 

\DeclareUnicodeCharacter{1E45}{\.{n}}
\DeclareUnicodeCharacter{1E41}{\.{m}}
\DeclareUnicodeCharacter{2003}{\quad}
\DeclareUnicodeCharacter{2009}{\thinspace}
\DeclareUnicodeCharacter{2002}{\enspace{}}
\DeclareUnicodeCharacter{2005}{\thinspace}
\DeclareUnicodeCharacter{0263}{\textipa{G}}
\DeclareUnicodeCharacter{A0}{~}
\DeclareUnicodeCharacter{2460}{\textcircled{\scriptsize{1}}}
\DeclareUnicodeCharacter{2461}{\textcircled{\scriptsize{2}}}
\DeclareUnicodeCharacter{2462}{\textcircled{\scriptsize{3}}}
\DeclareUnicodeCharacter{2463}{\textcircled{\scriptsize{4}}}
\DeclareUnicodeCharacter{2464}{\textcircled{\scriptsize{5}}}
\DeclareUnicodeCharacter{2465}{\textcircled{\scriptsize{6}}}
\DeclareUnicodeCharacter{2466}{\textcircled{\scriptsize{7}}}
\DeclareUnicodeCharacter{2467}{\textcircled{\scriptsize{8}}}
\DeclareUnicodeCharacter{2468}{\textcircled{\scriptsize{9}}}
\DeclareUnicodeCharacter{2070}{\textsuperscript{0}}
\DeclareUnicodeCharacter{2074}{\textsuperscript{4}}
\DeclareUnicodeCharacter{2075}{\textsuperscript{5}}
\DeclareUnicodeCharacter{2076}{\textsuperscript{6}}
\DeclareUnicodeCharacter{2077}{\textsuperscript{7}}
\DeclareUnicodeCharacter{2078}{\textsuperscript{8}}
\DeclareUnicodeCharacter{2079}{\textsuperscript{9}}
\DeclareUnicodeCharacter{02C2}{<}
\DeclareUnicodeCharacter{2033}{''}
\DeclareUnicodeCharacter{0229}{\c{e}}
\DeclareUnicodeCharacter{016F}{\r{u}}
\DeclareUnicodeCharacter{3AC}{\relax\ifmmode\acute{\alpha}\else $\acute{\alpha}$\fi}
\DeclareUnicodeCharacter{3AD}{\relax\ifmmode\acute{\varepsilon}\else $\acute{\varepsilon}$\fi}
\DeclareUnicodeCharacter{3AE}{\relax\ifmmode\acute{\eta}\else $\acute{\eta}$\fi}
\DeclareUnicodeCharacter{3AF}{\relax\ifmmode\acute{\iota}\else $\acute{\iota}$\fi}
\DeclareUnicodeCharacter{3CC}{\relax\ifmmode\acute{o}\else $\acute{o}$\fi}
\DeclareUnicodeCharacter{3CD}{\relax\ifmmode\acute{\upsilon}\else $\acute{\upsilon}$\fi}
\DeclareUnicodeCharacter{3CE}{\relax\ifmmode\acute{\omega}\else $\acute{\omega}$\fi}
\DeclareUnicodeCharacter{391}{A}
\DeclareUnicodeCharacter{392}{B}
\DeclareUnicodeCharacter{395}{E}
\DeclareUnicodeCharacter{396}{Z}
\DeclareUnicodeCharacter{397}{H}
\DeclareUnicodeCharacter{399}{I}
\DeclareUnicodeCharacter{39A}{K}
\DeclareUnicodeCharacter{39C}{M}
\DeclareUnicodeCharacter{39D}{N}
\DeclareUnicodeCharacter{39F}{O}
\DeclareUnicodeCharacter{3A1}{P}
\DeclareUnicodeCharacter{3A4}{T}
\DeclareUnicodeCharacter{3A7}{X}

\DeclareUnicodeCharacter{27E6}{\relax\ifmmode \llbracket \else $\llbracket$\fi}
\DeclareUnicodeCharacter{27E7}{\relax\ifmmode \rrbracket \else $\rrbracket$\fi}

\DeclareUnicodeCharacter{1D434}{\relax\ifmmode A \else $A$\fi}
\DeclareUnicodeCharacter{1D435}{\relax\ifmmode B \else $B$\fi}
\DeclareUnicodeCharacter{1D436}{\relax\ifmmode C \else $C$\fi}
\DeclareUnicodeCharacter{1D437}{\relax\ifmmode D \else $D$\fi}
\DeclareUnicodeCharacter{1D438}{\relax\ifmmode E \else $E$\fi}
\DeclareUnicodeCharacter{1D439}{\relax\ifmmode F \else $F$\fi}
\DeclareUnicodeCharacter{1D43A}{\relax\ifmmode G \else $G$\fi}
\DeclareUnicodeCharacter{1D43B}{\relax\ifmmode H \else $H$\fi}
\DeclareUnicodeCharacter{1D43C}{\relax\ifmmode I \else $I$\fi}
\DeclareUnicodeCharacter{1D43D}{\relax\ifmmode J \else $J$\fi}
\DeclareUnicodeCharacter{1D43E}{\relax\ifmmode K \else $K$\fi}
\DeclareUnicodeCharacter{1D43F}{\relax\ifmmode L \else $L$\fi}
\DeclareUnicodeCharacter{1D440}{\relax\ifmmode M \else $M$\fi}
\DeclareUnicodeCharacter{1D441}{\relax\ifmmode N \else $N$\fi}
\DeclareUnicodeCharacter{1D442}{\relax\ifmmode O \else $O$\fi}
\DeclareUnicodeCharacter{1D443}{\relax\ifmmode P \else $P$\fi}
\DeclareUnicodeCharacter{1D444}{\relax\ifmmode Q \else $Q$\fi}
\DeclareUnicodeCharacter{1D445}{\relax\ifmmode R \else $R$\fi}
\DeclareUnicodeCharacter{1D446}{\relax\ifmmode S \else $S$\fi}
\DeclareUnicodeCharacter{1D447}{\relax\ifmmode T \else $T$\fi}
\DeclareUnicodeCharacter{1D448}{\relax\ifmmode U \else $U$\fi}
\DeclareUnicodeCharacter{1D449}{\relax\ifmmode V \else $V$\fi}
\DeclareUnicodeCharacter{1D44A}{\relax\ifmmode W \else $W$\fi}
\DeclareUnicodeCharacter{1D44B}{\relax\ifmmode X \else $X$\fi}
\DeclareUnicodeCharacter{1D44C}{\relax\ifmmode Y \else $Y$\fi}
\DeclareUnicodeCharacter{1D44D}{\relax\ifmmode Z \else $Z$\fi}
\DeclareUnicodeCharacter{1D44E}{\relax\ifmmode a \else $a$\fi}
\DeclareUnicodeCharacter{1D44F}{\relax\ifmmode b \else $b$\fi}
\DeclareUnicodeCharacter{1D450}{\relax\ifmmode c \else $c$\fi}
\DeclareUnicodeCharacter{1D451}{\relax\ifmmode d \else $d$\fi}
\DeclareUnicodeCharacter{1D452}{\relax\ifmmode e \else $e$\fi}
\DeclareUnicodeCharacter{1D453}{\relax\ifmmode f \else $f$\fi}
\DeclareUnicodeCharacter{1D454}{\relax\ifmmode g \else $g$\fi}
\DeclareUnicodeCharacter{1D456}{\relax\ifmmode i \else $i$\fi}
\DeclareUnicodeCharacter{1D457}{\relax\ifmmode j \else $j$\fi}
\DeclareUnicodeCharacter{1D458}{\relax\ifmmode k \else $k$\fi}
\DeclareUnicodeCharacter{1D459}{\relax\ifmmode l \else $l$\fi}
\DeclareUnicodeCharacter{1D45A}{\relax\ifmmode m \else $m$\fi}
\DeclareUnicodeCharacter{1D45B}{\relax\ifmmode n \else $n$\fi}
\DeclareUnicodeCharacter{1D45C}{\relax\ifmmode o \else $o$\fi}
\DeclareUnicodeCharacter{1D45D}{\relax\ifmmode p \else $p$\fi}
\DeclareUnicodeCharacter{1D45E}{\relax\ifmmode q \else $q$\fi}
\DeclareUnicodeCharacter{1D45F}{\relax\ifmmode r \else $r$\fi}
\DeclareUnicodeCharacter{1D460}{\relax\ifmmode s \else $s$\fi}
\DeclareUnicodeCharacter{1D461}{\relax\ifmmode t \else $t$\fi}
\DeclareUnicodeCharacter{1D462}{\relax\ifmmode u \else $u$\fi}
\DeclareUnicodeCharacter{1D463}{\relax\ifmmode v \else $v$\fi}
\DeclareUnicodeCharacter{1D464}{\relax\ifmmode w \else $w$\fi}
\DeclareUnicodeCharacter{1D465}{\relax\ifmmode x \else $x$\fi}
\DeclareUnicodeCharacter{1D466}{\relax\ifmmode y \else $y$\fi}
\DeclareUnicodeCharacter{1D467}{\relax\ifmmode z \else $z$\fi}

\firstpage{1}

  \articlenumber{958}

\pubvolume{9}
\issuenum{9}
\pubyear{2021}
\copyrightyear{2021}
\externaleditor{Academic Editor: J. Tenreiro~Machado}

\datereceived{11 March 2021}
\dateaccepted{15 April 2021}
\datepublished{25 April 2021}
\hreflink{https://doi.org/\linebreak 10.3390/math9090958}

\usepackage[T5,T1]{fontenc}
\usepackage[russian,english]{babel}

\Title{Statistical Significance Revisited}

\TitleCitation{Statistical Significance Revisited}

\Author{Maike~Tormählen~\textsuperscript{1}, Galiya~Klinkova~\textsuperscript{2} and Michael~Grabinski~\textsuperscript{2}\textsuperscript{,}*}
\AuthorNames{Maike Tormählen, Galiya Klinkova and Michael Grabinski}

\AuthorCitation{Tormählen, M.; Klinkova, G.; Grabinski, M.}

\address{\textsuperscript{1} \quad Department of Information Management, Neu-Ulm University, Wileystr. 1, 89231 Neu-Ulm, Germany; research@tormaehlen.net

\textsuperscript{2} \quad Department of Business and Economics, Neu-Ulm University, Wileystr. 1, 89231 Neu-Ulm, Germany; galiya.klinkova@uni-neu-ulm.de}

\corres{Correspondence: michael.grabinski@uni-neu-ulm.de}

\abstract{Statistical significance measures the reliability of a result obtained from a random experiment. We investigate the number of repetitions needed for a statistical result to have a certain significance. In the first step, we consider binomially distributed variables in the example of medication testing with fixed placebo efficacy, asking how many experiments are needed in order to achieve a significance of 95\%. In the next step, we take the probability distribution of the placebo efficacy into account, which to the best of our knowledge has not been done so far. Depending on the specifics, we show that in order to obtain identical significance, it may be necessary to perform twice as many experiments than in a setting where the placebo distribution is neglected. We proceed by considering more general probability distributions and close with comments on some erroneous assumptions on probability distributions which lead, for instance, to a trivial explanation of the fat tail.}

\keyword{statistical significance; confidence; medication tests; central limit theorem; fat~tail}

\makeatletter
\DeclareRobustCommand*\textsubscript[1]{%
  \@textsubscript{\selectfont#1}}
\def\@textsubscript#1{%
  {\m@th\ensuremath{_{\mbox{\fontsize\sf@size\z@#1}}}}}
\makeatother

\usepackage{sansmath}

\begin{document}
\section{Introduction \label{sect:sec1-mathematics-1159857}}

Statistical results never hold with absolute certainty. In fact, every statement comes with a certain probability of being erroneous. Performing exactly the same experiment an infinite number of times (limit $\left. n\rightarrow\infty \right. $, where $n $ is the number of cases or experiments) can, in theory, lead to absolutely certain statements. However, in reality everything is finite, so that statistical statements can never be a hundred percent assured.

Hence, a reasonable question is whether a given finite number of experiments $n $ is considered large enough to yield a reliable result. The search for the probability of the result being correct leads to the concept of statistical significance.

In the field of experimental physics (e.g., particle colliders), the above mentioned $n $ can go up to magnitudes of $10^{23} $. This number is so large that the corresponding statistical significance is (almost) 100\%, cf. also~\cite{B1-mathematics-1159857,B2-mathematics-1159857}. Therefore, nobody honestly questions statistical significance in that context. However, in many other branches of research $10^{23} $ repetitions are simply impossible to perform.

In general, there are two types of information that can be investigated by use of statistical methods:

\begin{enumerate}[label=\arabic*.]
\item Quantities with a fixed value like a length or the number of viruses in a certain amount of blood.
\item Quantities with no reliable value like the share of people who like the color red better than green or who are immunized by vaccination.
\end{enumerate}

The first kind is the usual one investigated in experimental physics, where in most cases the error of an experiment can be easily determined, see e.g.,~\cite{B3-mathematics-1159857}. The second kind is tricky. Such quantities are usually measured indirectly (by retrieving a subjective perception instead of determining an objective number), and they are not defined accurately in a mathematical sense: what exactly does “\emph{I like red better than green}” mean? Is it the potential color of a car, a house, or a bouquet of flowers? What does \emph{like} mean? Does it mean the willingness to spend money for it, or an increase of the individual dopamine level?

This second kind of measurement is a typical subject to psychology and economics. A lot of tools for the analysis and interpretation of such experiments have been developed, starting almost a century ago~\cite{B4-mathematics-1159857} and still being discussed in current conference talks~\cite{B5-mathematics-1159857}. Nowadays, software is used more and more frequently in this context. The use of such readymade tools is risky, as---unless well understood---they are black boxes whose unquestioned use can easily lead to severe mistakes and misinterpretations.

As an example, one may consider the following problem. The new Covid-19 vaccine by Pfizer and BioNTech is supposed to lead to an immunity of 95\%. A test with only two probands could lead to possible outcomes of immunities of 0\%, 50\%, or 100\%. The corresponding probabilities are 0.25\%, 9.5\% and 90.25\%, respectively. This means: about a tenth of such experiments will lead to an efficacy of 50\%. In order to find more reliable results, a greater number of experiments has to be performed. This paper investigates how many experiments are actually needed.

As the number of experiments is always a natural number and probabilities are rational numbers, the sample of two probands can never show an immunity of exactly 95\%. In order to also avoid other complications, we will work with a continuum limit in \mbox{\sect{sect:sec2-mathematics-1159857}}. This is by no means new but rarely used in the discussion of binominal problems. As an example, we will calculate the statistical significance of the new vaccine.

While \sect{sect:sec2-mathematics-1159857} is essentially a repetition of the state of the art, in \sect{sect:sec3-mathematics-1159857} we will present completely new findings. There we will discuss bivariate distributions, assuming that the efficacy of the placebo in the control group comes with a certain distribution as well. We investigate whether, under this assumption, the number of experiments necessary for a certain significance changes. Up to our knowledge, this question has not been posed before. We find that the number of trials needed to obtain a certain statistical significance is way higher than generally suspected. One may need twice as many or even more probands than previously believed.

In \sect{sect:sec4-mathematics-1159857} we discuss more general probability distributions. Consider, for instance, the following example: we assume that 80\% of all men are taller than women. How many men and women must be tested to confirm this conjecture with a significance of 95\%? To answer this question, the height distribution of men and women has to be known. Even if the distributions are Gaussian, the question is difficult to answer in this general form. 

In \sect{sect:sec5-mathematics-1159857} we will discuss typical mistakes made in the context of statistical significance which are not so new. In particular, we discuss the problem of wrongly assuming a Gaussian distribution (\sect{sect:sec5dot1-mathematics-1159857}), taking as example an experiment which tries to prove that a court sentence is influenced by a person’s outer appearance~\cite{B6-mathematics-1159857}. The experiment claims a statistical significance of 95\%. However, due to a misuse of the central limit theorem, there is probably no significance at all. Furthermore, we look at situations where the assumption of a Gaussian distribution is justified but restricted to certain values \mbox{(\sect{sect:sec5dot2-mathematics-1159857}}). In finance, the return on investment may take a tremendously high value but cannot decrease by more than 100\%. Even if the returns show a perfect Gaussian cut off for values below 100\%, the standard deviation does not determine the width of the Gaussian distribution in this case. Ignoring this fact leads to a trivial explanation of the fat tail. In \sect{sect:sec5dot3-mathematics-1159857}, we discuss the typical requirement of a 95\% significance. Especially in clinical testing, severe mistakes can be made in this context. Repeating an experiment with this significance for 20 times with no effective change of the medication will statistically, by pure chance, lead to an outcome in which the medication is effective enough to be clinically approved. Closely related to this is the problem of false-positive selection discussed in \sect{sect:sec5dot4-mathematics-1159857}, one of the major reasons why a large number of publications based on heuristic studies is proven wrong later. 

In \sect{sect:sec6-mathematics-1159857} we will briefly discuss the main findings and draw conclusions.

\section{Binomial Distribution in the Continuum Limit: The One-Dimensional Case \label{sect:sec2-mathematics-1159857}}

We start by studying binomially distributed probabilities. The formulas presented in this section (cf.~\cite{B7-mathematics-1159857} for example) assume the independent measurement of a quantity that takes only two values like “bigger/smaller than” or “effective/non-effective”. As an explicit application, one may think of an efficacy test in pharmaceutical studies.

We define the following quantities:\begin{equation}
\label{eq:FD1-mathematics-1159857}
\begin{array}{l}
{M:{\ \text{number}\ \text{of}\ \text{experiments}}} \\
{q:{\ \text{efficacy}\ \text{or}\ \text{validity}\ \text{measured}\ \text{in}\ \text{these}\ }M{\ \text{experiments}\ }\left( 1 \right)} \\
\left. Q:{\ \text{true}\ \text{efficacy}\ \text{or}\ \text{validity},\ \text{measured}\ \text{for}\ }M\rightarrow\infty \right. \\
\end{array}
\tag{1}
\end{equation}
      with $M \in \mathbb{N} $ and $q,Q \in \left\lbrack {0,1} \right\rbrack \subset \mathbb{Q} $. The probability density $w\left( {q,Q,M} \right) $ for $q $ being measured in $M $ experiments with true efficacy $Q $ is given by
      \begin{equation}
\label{eq:FD4-mathematics-1159857}
\text{w}\left( {\text{q},\text{Q},\text{M}} \right) = \text{Q}^{\text{q}\text{$\cdotp $}\text{M}}\text{$\cdotp $}\left( {1 - \text{Q}} \right)^{{({1 - \text{q}})}\text{$\cdotp $}\text{M}}\text{$\cdotp $}\ \left( \begin{array}{c}
\text{M} \\
{\text{q}\text{$\cdotp $}\text{M}} \\
\end{array} \right).
\tag{2}
\end{equation}

In a medication test, $q $ and $Q $ are rational numbers, $q\text{$\cdotp $}M \in \mathbb{N} $ is the number of experiments with effective medication and $\left( {1 - q} \right)\text{$\cdotp $}M \in \mathbb{N} $ is the corresponding number of experiments with ineffective medication. For $k = q\text{$\cdotp $}M $ the binomial theorem yields the normalization condition:\begin{equation}
\label{eq:FD5-mathematics-1159857}
\sum\limits_{\text{k} = 0}^{\text{M}}\text{w}\left( {\frac{\text{k}}{\text{M}},\text{Q},\text{M}} \right) = 1.
\tag{3}
\end{equation}

Using (2), we can calculate probabilities for certain scenarios. For instance, let us consider the probability $W\left( {q,Q,M} \right) $ of measuring the validity $q $ to be greater than the true one (i.e., $Q $) within $M $ experiments:\begin{equation}
\label{eq:FD6-mathematics-1159857}
W\left( {q,Q,M} \right) = \sum\limits_{k = 1 + \lfloor q\text{$\cdotp $}M\rfloor}^{M}Q^{k}\text{$\cdotp $}\left( {1 - Q} \right)^{M - k}\text{$\cdotp $}\left( \begin{array}{c}
M \\
k \\
\end{array} \right)
\tag{4}
\end{equation}

\begin{figure}[H]
\widefigure
\includegraphics[scale=1.35]{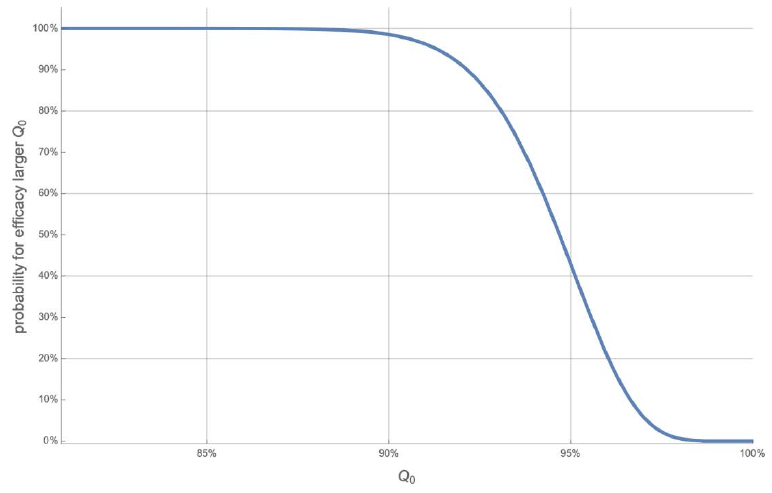}
\caption{Probability or significance for the BioNTech Pfizer vaccine showing an efficacy of Q\_0 or higher.}
\label{fig:mathematics-1159857-f001}
\end{figure}

Here, $\left\lfloor ~ \right\rfloor $ denotes the floor function. The floor $\left\lfloor q\text{$\cdotp $}M \right\rfloor $ is necessary as the summation index needs to be a natural number, and for arbitrary $q \in \left\lbrack {0,1} \right\rbrack $ (e.g., $q = 0.123 $) and $M $ (e.g., $M $ = 50) the product $q\text{$\cdotp $}M $ is likely not to be a positive integer. Equation (4) can be recast as
\vspace{6pt}
\end{paracol}
\nointerlineskip
  \begin{equation}
\label{eq:FD2-mathematics-1159857}
W\left( {q,Q,M} \right) = \frac{M!\left( {1 - Q} \right)^{M - \lfloor\text{q}\text{$\cdotp $}\text{M}\rfloor - 1}{Q^{\lfloor\text{q}\text{$\cdotp $}\text{M}\rfloor + 1}}_{2}F_{1}\left( {1,\left\lfloor \text{q}\text{$\cdotp $}\text{M} \right\rfloor + 1 - M;\left\lfloor \text{q}\text{$\cdotp $}\text{M} \right\rfloor + 2;\frac{Q}{Q - 1}} \right)}{\left( {M - \left\lfloor \text{q}\text{$\cdotp $}\text{M} \right\rfloor - 1} \right)!\left( {\left\lfloor \text{q}\text{$\cdotp $}\text{M} \right\rfloor + 1} \right)!},
\tag{5}
\end{equation}
\begin{paracol}{2}
\switchcolumn
\noindent      where \textsubscript{2}\emph{F}\textsubscript{1} is the hypergeometric function (cf.~\cite{B7-mathematics-1159857}). Solving such an expression for $M $ is pretty challenging---even numerically. Hence---and also for other reasons---introducing a continuum limit (see e.g.,~\cite{B8-mathematics-1159857} for further details) is helpful at this point.

To do so, we transform the sum into an integral, keeping the above definitions (1). The constraint $q,Q \in \left\lbrack {0,1} \right\rbrack $ still holds, but now we have $M \in \mathbb{R}^{+} $. Since
      \begin{equation}
\label{eq:FD7-mathematics-1159857}
\int\limits_{0}^{1}\text{d}q~w\left( {q,Q,M} \right) = \frac{1}{M}
\tag{6}
\end{equation}
      the integral form of (4) is given by
      \begin{equation}
\label{eq:FD8-mathematics-1159857}
W\left( {q,Q,M} \right) = M\text{$\cdotp $}\int\limits_{q}^{1}\text{d}\widetilde{q}~w\left( {\widetilde{q},Q,M} \right).
\tag{7}
\end{equation}

In general, this integral (7) can only be solved numerically but unlike Equation (5) it does not require the computational inconvenient floor function.

We demonstrate the use of this integral on the basis of two examples. The authors of article~\cite{B9-mathematics-1159857} conclude that humans recognize homosexuality in men with 61\% certainty only by looking at their faces. (The article actually intends to show that widely-used face recognition software recognizes homosexual men with a hit rate of up to 81\%. This result leads to ethical concerns, for homosexuality being seen as a (severe) crime in certain countries. Hence, the software and its algorithm have not been published) If it were impossible to decide about homosexuality from a facial image, the hit rate would be 50\%. Performing such an experiment $M $ times and measuring $q = 61/100 $ raises the question:

\textbf{\boldmath{How many experiments}} $\textit{\textbf{M}} $ \textbf{\boldmath{are required to show that human assessment outperforms mere guessing (}}$\textit{\textbf{Q}}_{\textbf{0}}~ = ~\textbf{1}/\textbf{2} $\textbf{\boldmath{) with a certain probability (significance)}} $\textit{\textbf{p}} $\textbf{\boldmath{?}}

This question corresponds to the problem of determining the number of probands needed in a pharmaceutical study to sufficiently prove a drug’s efficacy. In medication testing, the answer is usually determined without using a continuum limit and most likely computed by the aid of ready-made software.

Here, we will give a general formula to answer the above question. First, one has to apply the proper normalization. In analogy to (6) we have
      \begin{equation}
\label{eq:FD9-mathematics-1159857}
\int\limits_{0}^{1}\text{d}Q~w\left( {q,Q,M} \right) = \frac{1}{M + 1},
\tag{8}
\end{equation}
      leading to the following expression for the significance
      \vspace{6pt}
\end{paracol}
\nointerlineskip
   \begin{equation}
\label{eq:FD10-mathematics-1159857}
p\left( {q,M} \right) = \left( {M + 1} \right)\text{$\cdotp $}\int\limits_{Q_{0}}^{1}\text{d}Q~w\left( {q,Q,M} \right) = 1 - \left( \begin{array}{c}
M \\
{q\text{$\cdotp $}M} \\
\end{array} \right)\left( {M + 1} \right)~B_{Q_{0}}\left( {1 + Mq,1 + M - Mq} \right)
\tag{9}
\end{equation}
\begin{paracol}{2}
\switchcolumn

Here, $B_{Q_{0}} $ denotes the incomplete beta function defined as
      \begin{equation}
\label{eq:FD11-mathematics-1159857}
B_{Q_{0}}\left( {1 + Mq,1 + M\left( {1 - q} \right)} \right) = \int\limits_{0}^{Q_{0}}\text{d}t~t^{Mq}~\left( {1 - t} \right)^{M - Mq}.
\tag{10}
\end{equation}

Though (9) in combination with (10) appears to be rather inconvenient, it turns out that functions like $B_{Q_{0}} $ can easily be calculated with arbitrary precision. Inserting the values $Q_{0} = 0.5 $, $q = 0.61 $, and $p = 0.9 $ into (9), the equation can be (numerically) solved for $M $, yielding $M \approx 34.809 $. This implies the requirement for a minimum of 35 trials in order to prove human superiority in the recognition of homosexual faces over mere guessing with an accuracy of 90\%.

The second example has already been mentioned in the introduction. The pharmaceutical companies BioNTech and Pfizer examined their Covid-19 vaccine in a study with 43,000 participants. The participants were divided into a test group receiving the actual vaccine and a control (placebo) group. Eventually, 170 of all participants got infected with Covid-19. Eight of these participants were part of the test group, thus the remaining \mbox{162 infected} participants belonged to the control group. More details can be found in~\cite{B10-mathematics-1159857}. (Actually, nine people from the test group were infected, but one person had to be excluded due to previous illness as stated in~\cite{B10-mathematics-1159857}) Essentially we have the following setup: The groups were equally segmented (21,500 people each in the test and in the control group). As it would be unethical to infect all 43,000 probands on purpose, they have been observed over a certain period of time. From the reported result of the control group, we assume that on average 162 out of 21,500 people get infected over this fixed period. This leads to the observation that in the (actually vaccinated) test group $162 - 8 = 154 $ people were presumably immune, corresponding to an immunization rate of $154/162 \approx 0.95064 $ or roughly 95\%, which is a remarkably good result. However, the question remains: How significant is this result? The probability $p $ for an efficacy $Q_{0} $ equal to or higher than $q = 154/162 \approx 0.95064 $ measured within $M = 162 $ experiments is calculated from (9):\begin{equation}
\label{eq:FD12-mathematics-1159857}
p\left( {q,M} \right) = \left( {M + 1} \right)\text{$\cdotp $}\int\limits_{Q_{0}}^{1}\text{d}Q~w\left( {q,Q,M} \right)
\tag{11}
\end{equation}

This equation can be interpreted as a function $p\left( Q_{0} \right) $ with fixed $M = 162 $ and $q \approx 0.95 $. Inserting $Q_{0} = 0.95 $ leads to a probability $p \approx 0.429 $ or roughly 43\% which appears to be rather low. Nevertheless, a slight change of the argument, e.g., $Q_{0} = 0.90 $, already leads to a way higher significance of $p \approx 0.985 $. Hence, in this experiment the vaccination has an efficacy of 90\% with a significance of roughly 99\%. \fig{fig:mathematics-1159857-f001} shows the corresponding plot of significance over efficacy. In particular, we observe that a variation of $Q_{0} $ in the range between 90\% and 100\% results in a considerable change in significance, whereas a variation of $Q_{0} $ in the range below 90\% only leads to relatively small changes in $p $. This is a typical behavior of binomially distributed quantities.      
      \end{paracol}

\begin{figure}[H]
\widefigure
\includegraphics[scale=1.5]{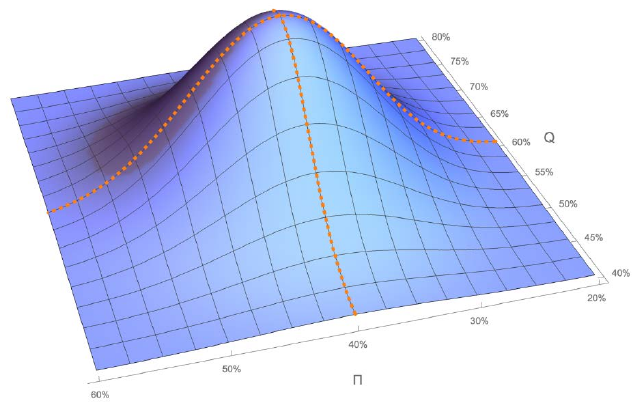}
\caption{Plot of
		  $w\left( {\frac{6}{10},Q,40} \right)\text{$\cdotp $}\omega\left( {\frac{4}{10},\Pi,40} \right) $.}
\label{fig:mathematics-1159857-f002}
\end{figure}

      \begin{paracol}{2}
      \switchcolumn

The result illustrated in \fig{fig:mathematics-1159857-f001} is possibly unknown to BioNTech or Pfizer. Usually, medication testing requires the achievement of a certain significance (mostly 95\%) rather than a given value $Q_{0} $, i.e., the medication must be proven to be more effective than the placebo. In our example, we assumed the placebo to have zero efficacy. However, in order to obtain more reliable results, a second control group, given neither a vaccine nor a placebo, should have been included in the experiment. 

The question of how to include the distribution of placebo efficacies into the underlying statistical model is our central object of investigation and will be discussed in the \mbox{next section.}

\section{Bivariate Binomial Distribution: The Two-Dimensional Case \label{sect:sec3-mathematics-1159857}}

So far, we have been assuming a fixed efficacy (zero in the example of BioNTech and Pfizer), in the placebo or control group. This is probably not the case in reality. More likely, the placebo efficacy is distributed in the same way as the medication efficacy, described as in Equation (2) with
      \begin{equation}
\label{eq:FD13-mathematics-1159857}
\begin{array}{l}
{N:~{\text{number}\ \text{of}\ \text{experiments}\ \text{in}\ \text{the}\ \text{control}\ \text{group}}} \\
{\pi:~{\text{efficacy}\ \text{or}\ \text{validity}\ \text{measured}\ \text{in}\ \text{these}}~N~\text{experiments}} \\
\left. \Pi:~{\text{true}\ \text{efficacy}\ \text{or}\ \text{validity},\ \text{measured}\ \text{for}\ }N\rightarrow\infty: \right. \\
\end{array}
\tag{12}
\end{equation}
\begin{equation}
\label{eq:FD16-mathematics-1159857}
\omega\left( {\pi,\Pi,N} \right) = \Pi^{\pi\text{$\cdotp $}N}\text{$\cdotp $}\left( {1 - \Pi} \right)^{{({1 - \pi})}\text{$\cdotp $}N}\text{$\cdotp $}\left( \begin{array}{c}
N \\
{\pi\text{$\cdotp $}N} \\
\end{array} \right).
\tag{13}
\end{equation}

The density $\omega\left( {\pi,\Pi,N} \right) $ describes the probability to measure an efficacy $\pi $ in the control group within $N $ trials when $\Pi $ is the true efficacy. Using the continuum limit as before, we have $\pi,\Pi \in \left\lbrack {0,1} \right\rbrack $ and  $N \in \mathbb{R}^{+} $.

The total probability density is given by $w\left( {q,Q,M} \right)\text{$\cdotp $}\omega\left( {\pi,\Pi,N} \right) $ and has been plotted in \fig{fig:mathematics-1159857-f002}. It is the two-dimensional version of a binomial distribution in the continuum limit. After normalization, the volume covered by the graph in \fig{fig:mathematics-1159857-f002} is one. This probability density will serve as a basis for discussing a question similar to the one in \sect{sect:sec2-mathematics-1159857}:
\end{paracol}
\nointerlineskip

\begin{paracol}{2}
\switchcolumn

\textbf{\boldmath{How many experiments are needed to assure with a certain significance that a medication is better than a placebo?}}

In a typical placebo-controlled medication study, the parameters $q,\pi,M,N $ are known. A natural question is whether the true efficacy $Q $ is greater than or equal to the placebo efficacy $\Pi $ plus some buffer, i.e.,
      \begin{equation}
\label{eq:FD17-mathematics-1159857}
Q \geq \Pi + \delta.
\tag{14}
\end{equation}

The buffer $\delta $ could have been included in the consideration of the previous section for more general results but was left out for the sake of simplicity there. In medical research, we normally have $\delta = 0 $ and often $M = N, $ but we want to keep it general here. The probability $p\left( {q,\pi,M,N} \right) $ that the true efficacy of the medication is by a margin of \emph{$\delta$} above the efficacy of the control group is then given by the following equation, where (14) serves to determine the integration boundary for the second integrand:\begin{equation}
\label{eq:FD18-mathematics-1159857}
p\left( {q,\pi,M,N,\delta} \right) = \left( {M + 1} \right)\left( {N + 1} \right)\text{$\cdotp $}\int\limits_{\delta}^{1}\text{d}Q~\int\limits_{0}^{Q - \delta}{\text{d}\Pi\ }w\left( {q,Q,M} \right)\text{$\cdotp $}\omega\left( {\pi,\Pi,N} \right)
\tag{15}
\end{equation}

The factor $\left( {M + 1} \right)\left( {N + 1} \right) $ is necessary for normalization, cf. (8). The densities $w\left( {q,Q,M} \right) $ and $\omega\left( {\pi,\Pi,N} \right) $ can be specified by inserting the expressions (2) and (13). Equation (15) cannot be solved analytically but simplified as follows:

\end{paracol}
\nointerlineskip
\begin{equation}
\label{eq:FD19-mathematics-1159857}
p\left( {q,~\pi,~M,~N,\delta} \right) = \left( {M + 1} \right)\left( {N + 1} \right)\left( \begin{array}{c}
M \\
{q\text{$\cdotp $}M} \\
\end{array} \right)\left( \begin{array}{c}
N \\
{\pi\text{$\cdotp $}N} \\
\end{array} \right)\int\limits_{\delta}^{1}\text{d}Q~\left( {1 - Q} \right)^{M - Mq}\text{$\cdotp $}Q^{Mq}\text{$\cdotp $}B_{Q - \delta}\left( {N\pi + 1,N - N\pi + 1} \right),
\tag{16}
\end{equation}
\begin{paracol}{2}
\switchcolumn

 \noindent    \textls[-25]{ where  $B_{Q - \delta} $ is the incomplete beta function as defined in (10). Determining $p\left( {q,\ \pi,\ M,\ N,\delta} \right) $} explicitly takes a high computational effort but can be simplified in the particular case of $N\pi,\ Mq \in \mathbb{N} $. As can be shown by the use of partial integration in (10) and complete induction, the incomplete beta function takes the form'
     \vspace{6pt}

\end{paracol}
\nointerlineskip
      \begin{equation}
\label{eq:FD20-mathematics-1159857}
B_{Q - \delta}\left( {N\pi + 1,N - N\pi + 1} \right) = \sum\limits_{k = 1 + N\text{$\cdotp $}\pi}^{N}\frac{\left( {N\text{$\cdotp $}\pi} \right)!\text{$\cdotp $}\left( {N - N\text{$\cdotp $}\pi} \right)!}{k!\text{$\cdotp $}\left( {N - k} \right)!}\ \left( {Q - \delta} \right)^{k}\ \left( {1 - Q + \delta} \right)^{N - k}.
\tag{17}
\end{equation}
\begin{paracol}{2}
\switchcolumn

Inserting this into Equation (16) yields

\end{paracol}
\nointerlineskip
     \begin{equation}
\label{eq:FD21-mathematics-1159857}
p\left( {q,~\pi,~M,~N,\delta} \right) = \frac{\left( {M + 1} \right)!\left( {N + 1} \right)!}{\left( {Mq} \right)!\left( {M - Mq} \right)!}\sum\limits_{k = 1 + N\text{$\cdotp $}\pi}^{N}\frac{1}{k!\text{$\cdotp $}\left( {N - k} \right)!}~\int\limits_{\delta}^{1}\text{d}Q~\left( {1 - Q} \right)^{M - Mq}~Q^{Mq}\left( {Q - \delta} \right)^{k}~\left( {1 - Q + \delta} \right)^{N - k}.~
\tag{18}
\end{equation}
\begin{paracol}{2}
\switchcolumn

With $Mq \in \mathbb{N} $, the integral in (18) contains solely integer exponents and therefore turns into a polynomial in $Q $ which can be easily solved by use of a computer algebra system like Mathematica. Whenever $N\pi,~Mq \in \mathbb{N} $ does not hold---as is the case in the following example ---the more complicated expressions (15) or (16) have to be used for explicit computations.

A nonvanishing and not uniformly distributed placebo efficacy actually has a severe impact. For a better understanding, let us consider the following exemplary medication test: A medication shows efficacy of 45\% ($q = \frac{45}{100} $) while the corresponding placebo has an efficacy of 40\% ($\pi = \frac{4}{10} $). $M = N $ is assumed for simplicity. How many probands are needed in each group to probe with a significance of 99\% that the medication is at least more effective than the placebo? In the first step, let us ignore the efficacy distribution of the placebo and use Equation (11) as in \sect{sect:sec2-mathematics-1159857}. This yields:\begin{equation}
\label{eq:FD22-mathematics-1159857}
\left. p\left( {\frac{45}{100},~M} \right) = \left( {M + 1} \right)\text{$\cdotp $}\int\limits_{4/10}^{1}\text{d}Q~w\left( {\frac{45}{100},Q,M} \right) \geq \frac{99}{100}\text{\quad\quad}\Rightarrow\text{\quad\quad}\textit{\textbf{M}} \geq \textbf{522} \right.
\tag{19}
\end{equation}

Now we use Equation (15) with $N = M $ in order to take the placebo efficacy into account:
\end{paracol}
\nointerlineskip
\begin{equation}
\label{eq:FD23-mathematics-1159857}
\left. p\left( {\frac{45}{100},\frac{4}{10},~M,~M,0} \right) = \left( {M + 1} \right)^{2}\text{$\cdotp $}\int\limits_{0}^{1}\text{d}Q~\int\limits_{0}^{Q}\text{d}\Pi\ w\left( {\frac{45}{100},Q,M} \right)\text{$\cdotp $}\omega\left( {\frac{40}{100},\Pi,M} \right) \geq \frac{99}{100}~~~~\Rightarrow~~\textit{\textbf{M}} \geq \textbf{1059} \right.
\tag{20}
\end{equation}
\begin{paracol}{2}
\switchcolumn

\textls[-35]{(Equation (20) can be solved analytically, but this is rather tedious. $p\left( {\frac{45}{100},\frac{4}{10},\ 1059,1059,0} \right)  $ }$ \approx 0.990009678 $ and $p\left( {\frac{45}{100},\frac{4}{10},\ 1058,1058,0} \right)\approx 0.989980349 $ can be computed by use of hypergeometric functions. Using Mathematica, the integration takes about 40 h CPU time each. The result is a rational number with almost $10^{9000} $ digits in its numerator and denominator. A numerical solution cannot be achieved with Mathematica, as the numerical error cannot be exactly determined. For a good estimation of the result, one can use the single integral (16) and split the integration interval into evenly spaced subintervals. A numerical integration by, say, 10,000 intervals leads (surprisingly fast, in less than a second) to very accurate results).

This calculation shows that by taking the distribution of the placebo efficacy into account, more than twice as many probands are needed in order to achieve the same significance. This result suggests that in many pharmaceutical studies the actual statistical significance may be way lower than specified, simply because the distribution of placebo efficacy is not taken into account in the computation. Especially in cases where the placebo efficacy is close to the medication efficacy, the difference will be rather significant.

\section{What Happens for Arbitrary Probability Distributions? \label{sect:sec4-mathematics-1159857}}

So far, we have restricted our considerations to binomially distributed quantities. In general, random variables can be associated with other probability distributions. For instance, the returns of stock portfolios or the body height of men and women take continuous values. For any quantitative statement, the corresponding probability distributions have to be known. One often (erroneously) assumes a Gaussian distribution, which will be commented on in \sect{sect:sec5dot1-mathematics-1159857}. (See~\cite{B11-mathematics-1159857} for a comment on how to avoid common mistakes when determining a distribution).

Let us assume two samples, for instance, a group of women and a group of men. Their respective body heights are given by two variables $x_{1} $ and $x_{2} $, statistically described by the corresponding probability distributions $p_{1}\left( x_{1} \right) $ and $p_{2}\left( x_{2} \right) $. We now ask for the probability that a subject from sample 2 with height $x_{2} $ is taller than an arbitrary subject from sample 1 plus some buffer, i.e., $x_{2} \geq x_{1} + \delta $. In analogy to Equation (15), the corresponding probability $p\left( \delta \right) $ is given by
      \begin{equation}
\label{eq:FD24-mathematics-1159857}
p\left( \delta \right) = \int\limits_{- \infty}^{\infty}\text{d}x_{2}~\int\limits_{- \infty}^{x_{2} - \delta}\text{d}x_{1}~p_{1}\left( x_{1} \right)\text{$\cdotp $}p_{2}\left( x_{2} \right).
\tag{21}
\end{equation}

Without explicit knowledge of the distributions $p_{1}\left( x_{1} \right) $ and $p_{2}\left( x_{2} \right) $, no further conclusion is possible. Assuming Gaussian distributions, the integral over $x_{1} $ can be rewritten by use of an error function as
      \begin{equation}
\label{eq:FD25-mathematics-1159857}
\int\limits_{- \infty}^{x_{2} - \delta}\text{d}x_{1}~\frac{1}{\sigma_{1}\sqrt{2\pi}}~e^{- \frac{1}{2}{(\frac{x_{1} - \mu_{1}}{\sigma_{1}})}^{2}} = \frac{1}{2} - \frac{1}{2}\text{erf}\left( \frac{x_{2} - \delta - \mu_{1}}{\sigma_{1}\sqrt{2}} \right).
\tag{22}
\end{equation}

Let us consider the case $p_{1}\left( x_{1} \right) = p_{2}\left( x_{2} \right) $ and $\delta = 0 $. In our example, this is the probability that from two randomly chosen men (or women) the first one is taller than the second one or vice versa. Within any symmetric distribution (not only a Gaussian one) this probability must be $1/2 $. For a Gaussian distribution, this is easy to verify by evaluating both integrals in (21). Needless to say that a single integral over the two distributions
      \begin{equation}
\label{eq:FD26-mathematics-1159857}
\int\limits_{- \infty}^{\infty}\text{d}x~\frac{1}{\sigma\sqrt{2\pi}}~e^{- \frac{1}{2}{(\frac{x - \mu}{\sigma})}^{2}}\text{$\cdotp $}\frac{1}{\sigma\sqrt{2\pi}}~e^{- \frac{1}{2}{(\frac{x - \mu}{\sigma})}^{2}} = \frac{1}{2\sigma\sqrt{\pi}}
\tag{23}
\end{equation}
      does not yield a probability. Depending on $\sigma $, this expression can take any real value and comes with the same dimension as $1/x $ ($1/\text{length} $ in our example), whereas probabilities are dimensionless by definition.

So far this discussion tackles distributions with one continuous variable and does not address statistical significance. As mentioned above, an explicit knowledge of the distributions $p_{1}\left( x_{1} \right) $ and $p_{2}\left( x_{2} \right) $ is needed in order to calculate probabilities. Determining a distribution with absolute certainty requires an infinite number of experiments, in which case the statistical significance for the probability calculated by use of (21) is trivially 100\%. Probing statistical significance for experiments with a continuous variable is devilishly complicated though well defined. In order not to be too abstract we will consider \mbox{an example.}

The intelligence quotient (IQ) is Gaussian distributed (for more details see e.g.,~\cite{B11-mathematics-1159857}). In Germany and many other industrialized countries, the average IQ is (approximately) 100. We assume that somebody wants to prove that in a certain area people are mentally impaired because they have an average IQ of 90. As we take the average IQ in Germany for granted, this question leads to the one-dimensional case of \sect{sect:sec2-mathematics-1159857}: How many people must be examined from the area in question in order to prove their lack of intellectual capability with a certain statistical significance?

Just probing one person yields a probability not so far below 50\% to find an IQ of \mbox{90 even} if the average IQ is 100 in this area. (The exact probability can be easily calculated) Taking two people already leads to a complicated computation. Without making a huge mistake, we can assume that the IQ of these two people always lies between 50 and 150 and that one can only measure integer IQ points, i.e., an IQ between 94.5 and 95.5 is counted as 95. As a consequence, testing only two people in Germany already gives a whole range of possibilities to find an average IQ of 90: 50 and 130, 51 and 129, and so on. All these pairs have different probabilities.

A generalization is straightforward but tedious. One can define $r_{n}\left( \left\langle {IQ} \right\rangle \right) $ as the probability to measure an average IQ of $\left\langle IQ \right\rangle $ from $n $ people in Germany. $r_{n}\left( \left\langle {IQ} \right\rangle \right) $ can be determined from the known distribution of IQs in Germany, but this turns out to be complex, as now \emph{n}-tuples rather than pairs contribute in the computation. The probabilities $r_{n}\left( \left\langle {IQ} \right\rangle \right) $ lead to a distribution ${\rho}_{n}\left( \left\langle {IQ} \right\rangle \right) $. Due to the central limit theorem, this distribution is Gaussian, peaked at 100 or at whatever average the original distribution of IQs comes with. For $n = 1 $ the distribution resembles the original IQ distribution. For higher $n $, it becomes sharper. The distribution ${\rho}_{n}\left( \left\langle {IQ} \right\rangle \right) $ is the equivalent of the probability $w\left( {q,Q,M} \right) $ in (2). For instance, one can use it to determine the probability $W $ to measure an average IQ of $90 $ or below (by integrating form $- \infty $ to $90 $) whereas $100 $ is the true average. $1 - W $ is the then the statistical significance that the average IQ of a subgroup is below $100 $ when $90 $ has be measured.

Note that there is a mathematical difference between $w\left( {q,Q,M} \right) $ and ${\rho}_{n}\left( \left\langle {IQ} \right\rangle \right) $. $M $ corresponds to $n $ and $q $ to $\left\langle IQ \right\rangle $, but $Q $ has no direct equivalent as it corresponds to the original distribution. ${\rho}_{n}\left( \left\langle {IQ} \right\rangle \right) $ is a \emph{functional} of this original distribution, in our example the IQ distribution in Germany. This illustrates that the derivation of closed formula for ${\rho}_{n}\left( \left\langle {IQ} \right\rangle \right) $ is intricate~\cite{B12-mathematics-1159857}.

\section{Other Shortcomings \label{sect:sec5-mathematics-1159857}}

In this section, we will comment on some common mistakes which are by no means new. As we observe them quite frequently, it appears appropriate to mention them in \mbox{this context.}

\subsection{Wrongly Assuming a Gaussian Distribution \label{sect:sec5dot1-mathematics-1159857}}

Assuming random variables to be Gaussian distributed simplifies all considerations significantly. In particular, determining the mean and standard deviation of experimental data becomes very easy. (The problem described in~\cite{B8-mathematics-1159857} has to be kept in mind here) Deriving or constructing other distributions is quite complicated in general, cf.~\cite{B11-mathematics-1159857}. But simplicity must not be the sole justification to work with Gaussian distributions by default. This problem has already been mentioned in centuries-old textbooks like~\cite{B4-mathematics-1159857}, “\emph{Everybody believes in the exponential law} [i.e., Gaussian distribution] \emph{of errors: the experimenters, because they think it can be proved by mathematics; and the mathematicians, because they believe it has been established by observation}”.

Today, the kind of distribution that best fits measured data can be automatically tested with suitable software, but this black-box approach can yield completely wrong results. To carve out this point, we start with a gedanken experiment considering a roulette game. Then we will comment on an experiment from psychology~\cite{B6-mathematics-1159857} where the same reasoning leads to completely wrong results, though the mistake is less obvious than in the \mbox{roulette experiment.}

Simulating one million roulette spins leads to an almost perfect uniform distribution from 0 to 36. The average of this distribution is approximately 18 (17.984 in our experiment), the standard deviation is approximately 10.4. The corresponding Gaussian distribution with $\mu = 17.984 $ and $\sigma = 10.4 $ will identify 18 as “\emph{the best number to bet on}” which is obviously a naïve conclusion as all numbers will actually appear with the same probability.

It seems unlikely that anyone will be deceived by this simple example, and data should be tested anyway before taking a distribution for granted. Unfortunately, this last step is quite frequently omitted in reality. Note that even in this example, a Gaussian distribution running from $- \infty $ to $+ \infty $ actually approximates the uniform distribution with a height of about 27,071 from 0 to 36 and 0 outside not too badly. However, in a roulette game numbers below 0 or above 36 do not have a finite probability, they are impossible. This problem is discussed in \sect{sect:sec5dot2-mathematics-1159857}.

The roulette experiment can be performed in another way. Instead of simulating or observing one million spins, the experimenter may call 100 casinos and ask for their individual averages within 1,000 spins. These 100 numbers show a narrow distribution around the value 18, which is almost perfectly Gaussian because of the central limit theorem. In our original experiment, we find an average of 17.990 and a standard deviation of 0.3. In fact, the two experiments do not address the same problem. The mistake lies in confusing the distribution of roulette numbers with the distribution of the corresponding averages. While the first one is a uniform distribution, the second one is always Gaussian.

Does this kind of confusion occur in real world experiments? The answer is yes. Normally it is done in a more subtle way. As an example, take the experiment described in~\cite{B6-mathematics-1159857} and its summary (and promotion) in~\cite{B13-mathematics-1159857}. In the experiment, 742 convicted murderers were rated by the trustworthiness of their faces. A total of 371 of the 742 candidates had been sentenced to death, whereas another 371 candidates had been sentenced for life imprisonment. On a scale from 1 (no trust) to 8 (very trustworthy), the first group was rated with a trustworthiness of 2.76, while the second group (life imprisonment) received an average score of 2.87. Though the difference is tiny, the authors claim a statistical significance of 95\%.

This result appears disconcerting, even without knowing the details of the underlying study. The faces were rated by humans on a scale from 1 to 8. Presumably, humans already have difficulties expressing a \emph{feeling} of trustworthiness in exactly eight categories. Hence, the assumption of an accuracy of $\pm 0.5 $ appears bold, as a difference of $2.87 - 2.76 = 0.11 $ can hardly be measured with whatever statistical procedure.

In~\cite{B6-mathematics-1159857}, the authors do not use (21) but an established software to calculate the resulting significance of 95\%. We assume that the software works correctly. The surprisingly high significance arises from the design of the experiment. The 742 candidates were rated by 208 volunteers in a complicated arrangement. The main point is that each face was rated around 30 times and only the average of these measurements was recorded and published~\cite{B14-mathematics-1159857}. (This arrangement was necessary in order not to compare apples and oranges. The candidates were split into subgroups, for instance distinguishing African Americans and Caucasians. Furthermore, not all 208 volunteers could rate all 742 faces. As a result, each face was rated about 30 times with varying number due to practical reasons. All in all, the experiment was performed very carefully, and there is no hint for mistakes) As with the roulette spins, such averages show a Gaussian distribution, even though not a perfect one because the number of repetitions was too small. Again, it is a direct result of the central limit theorem. Hence, we have a Gaussian-like distribution of the averages (each from about 30 measurements) for 371 candidates sentenced to death and the same number for the candidates sentenced to life imprisonment. The same procedure with a number of volunteers ten times higher would yield two distributions of averages from \mbox{300 measurements,} crucially increasing the significance. With a sufficient number of volunteers, one could reach any significance in this experiment.

As in the roulette example, instead of investigating the distributions of the original ratings, the authors of~\cite{B6-mathematics-1159857} examined the distribution of averages, which is always Gaussian and that allows for no conclusion on the distribution of the original ratings. Therefore, judging about statistical significance is impossible here. Because the original measurement has an accuracy of at most $\pm 0.5 $, measured values of $2.76 $ or $2.87 $ are identical and should be rounded to 3. The result of~\cite{B6-mathematics-1159857} should be stated as follows: Convicted murderers with life sentence or death role show an identical trustworthiness of 3 within the accuracy of \mbox{the experiment~\cite{B6-mathematics-1159857}.}

It appears appropriate to close this section with a rather cynical statement attributed to Carl Friedrich Gau{\ss} (*1777 - $\dagger$1855): \emph{Durch nichts wird mathematisches Unvermögen deutlicher als durch übergro{\ss}e Genauigkeit im Zahlenrechnen}. (English: \emph{Nothing proves mathematical incapacity better than too much accuracy in calculations}).

\subsection{Dealing with Parts of a Gaussian \label{sect:sec5dot2-mathematics-1159857}}

Some probability distributions, as for instance household incomes or stock returns, show a shape close to a Gaussian distribution. (For a more recent, detailed discussion of why portfolio theory is defective see e.g.,~\cite{B11-mathematics-1159857,B15-mathematics-1159857,B16-mathematics-1159857,B17-mathematics-1159857}) However, they behave slightly different, which has to be taken into account. These distributions, firstly, do not range from $- \infty $ to $+ \infty $, as there are no negative household incomes---especially when net incomes are considered. Stock returns can yield $- 100\% $ at minimum, but in theory they are not limited in the positive direction. For the distributions considered in~\cite{B18-mathematics-1159857,B19-mathematics-1159857}, negative returns are even far away from a total loss. Secondly, both household incomes and stock returns show a fat tail.

Especially in finance, the average return and standard deviation of a distributed quantity are used to determine $\mu $ and \emph{$\sigma$}, and hence the Gaussian distribution
        \begin{equation}
\label{eq:FD27-mathematics-1159857}
g\left( x \right) = \frac{1}{\sigma\sqrt{2\pi}}\text{$\cdotp $}e^{- \frac{{({x - \mu})}^{2}}{2\sigma^{2}}}.
\tag{24}
\end{equation}

Even if assuming a Gaussian distribution is completely justified, one has to take into account that it does not range from $- \infty $ to $+ \infty $ but rather starts at $x = 0 $.

Even for such distributions, i.e., distributions only defined on a subset of $\mathbb{R} $, one can calculate $\mu $ from the mean $m $ and $\sigma $ from the standard deviation $s $ by solving the following coupled equations, as shown in~\cite{B11-mathematics-1159857}, numerically:\begin{equation}
\label{eq:FD28-mathematics-1159857}
m\left( {\mu,\sigma} \right) = \frac{\mu\left( {2 - \Gamma_{r}\left( {- \frac{1}{2},\frac{\mu^{2}}{2\sigma^{2}}} \right)} \right)}{1 + \text{erf}\left( \frac{\mu}{\sqrt{2}\sigma} \right)}
\tag{25}
\end{equation}
\begin{equation}
\label{eq:FD29-mathematics-1159857}
\begin{array}{rl}
{s\left( {\mu,\sigma} \right)^{2}} & {= \frac{1}{\left( {1 + \text{erf}\left( \frac{\mu}{\sqrt{2}\sigma} \right)} \right)^{3}}(e^{- \frac{\mu^{2}}{2\sigma^{2}}}\sqrt{\frac{2}{\pi}}\mu\sigma\left( {1 + \text{erf}\left( \frac{\mu}{\sqrt{2}\sigma} \right)} \right)^{2}} \vspace{6pt}\\
 & {+ 2\left( \mu^{2} + \sigma^{2} \right.}\vspace{6pt} \\
 & {+ \text{erf}\left( \frac{\mu}{\sqrt{2}\sigma} \right)\left( 2\left( {\sigma - \mu} \right)\left( {\mu + \sigma} \right) + \left( {\mu^{2} + \sigma^{2}} \right)\text{erf}\left( \frac{\mu}{\sqrt{2}\sigma} \right) \right.} \vspace{6pt}\\
 & {\left. + \mu^{2}\left( {4 - \Gamma_{r}\left( {- \frac{1}{2},\frac{\mu^{2}}{2\sigma^{2}}} \right)} \right)\Gamma_{r}\left( {- \frac{1}{2},\frac{\mu^{2}}{2\sigma^{2}}} \right) \right))}\vspace{6pt} \\
 & {- \left( 5\mu^{2} + \sigma^{2} + \left( {\mu^{2} + \sigma^{2}} \right)\text{erf}\left( \frac{\mu}{\sqrt{2}\sigma} \right)\left( {2 + \text{erf}\left( \frac{\mu}{\sqrt{2}\sigma} \right)} \right) - 4\mu^{2}\Gamma_{r}\left( {- \frac{1}{2},\frac{\mu^{2}}{2\sigma^{2}}} \right) \right.} \vspace{6pt}\\
 & {\left. + \mu^{2}\Gamma_{r}\left( {- \frac{1}{2},\frac{\mu^{2}}{2\sigma^{2}}} \right)^{2} \right)\Gamma_{r}{\left( {- \frac{1}{2},\frac{\mu^{2}}{2\sigma^{2}}} \right))}} \\
\end{array}
\tag{26}
\end{equation}
		For a definition of $\text{erf} $ and $\Gamma_{r} $ see e.g.,~\cite{B11-mathematics-1159857}.

In finance, we often find that the wrong approach of $\mu = m $ and $\sigma = s $ is used. It is easy to show that this always leads to a $\sigma $ larger than the correct one calculated from (25) and (26). A too large $\sigma $ leads to nothing else but a fat tail. The severity of this mistake depends on how much of the distribution is cut off on the left-hand side. It is therefore important to consider not only the shape of a distribution but also the domain of $x $ in order to determine reasonable results for the specific context.

\subsection{The Dogma of 95\% Confidence \label{sect:sec5dot3-mathematics-1159857}}

A statistical significance of 95\% means that you have a chance of one out of twenty for your results to be pure coincidence. Most countries require this level of significance in medication tests, as it presents a compromise between efficacy and cost. However, it must be emphasized, that there is no intrinsic mathematical justification for this specific level of significance. Accepting every 20th medication to be potentially ineffective, simply limits the cost of new drugs.

A significance of 95\% should in general not be taken for granted. In order to find a suitable significance level, we must bear in mind that there are two possible interpretations which must not be confused. In \sect{sect:sec2-mathematics-1159857} we have seen that the BioNTech Pfizer vaccine is 95\% effective with a probability of 43\%. We may say that the vaccine immunizes 95\% of the probands, but we are only 43\% confident about this high rate. We may also say that we have a chance of 57\% that the vaccination is ineffective. The first interpretation sounds fantastic, the second one is disappointing at best. We have to keep in mind that the tuple “\mbox{efficacy = 95\%}, significance = 43\%” represents only one point of the curve in \fig{fig:mathematics-1159857-f001}. Considering the entire plot instead, we find that an efficacy of, say, 80\% comes with a significance of almost 100\%. This excludes the second interpretation, stating the ineffectiveness of the vaccine with a probability of 57\%. Illustrations of this kind are unfortunately rarely published, leaving us with the above ambiguity in the interpretation of the published numbers.

In~\cite{B6-mathematics-1159857} the authors intend to prove that court decisions depend on the defendant’s outer appearance. Although, as discussed earlier, this result appears to be highly questionable. Let us therefore deepen the discussion of this example here and suppose the experiment yields a confidence of 95\%, or, in other words, that every 20th experiment is meaningless. Again, this can mean that there is no relation of facial appearance and court decision and that the published result belongs to the 5\% of experiments that yield some relation by coincidence. It could also mean that there really is a relation, i.e., that a majority of judges are biased by facial appearance. While the first result would have no scientific impact, the second outcome would be quite valuable. Note that the same interpretation would be possible with a statistical significance of 50\%. A result with such a low significance would never have been published though it would still reveal a scandalous 50\% of judges sentencing by personal preference. Of course, by repeating the experiment several times with the same faces but different volunteers or different cases from different courts with the same volunteers would reveal which of the interpretations is the correct one.

Let us return to medication tests. If a clinical trial shows that the medication is more effective than the placebo with a significance of at least 95\%, the process takes an important step towards registration. (There is no requirement for an absolute minimal efficacy of the medication compared to the placebo) Clinical trials are often time-consuming and expensive, mainly due to the fact that the first survey does not necessarily lead to a sufficient level of efficacy within a significance of at least 95\%. In order to achieve the required efficacy and significance, it is possible that a medication takes around 50 attempts over ten years.

However, even if one compares a placebo to another placebo, every 20th test will statistically show sufficient efficacy when a significance of 95\% is assumed. This problem is particularly relevant if the efficacy of the medication is hard to measure and/or the placebo efficacy is high. A detailed publication of the mandatory test results would help to put the (most likely) correct efficacy of a medication into context. In fact, within five years after completion, less than half of the results of all medication tests in the United States are published, as  \fig{fig:mathematics-1159857-f003} shows. Even the two simple but important numbers for the efficacy of a medication and the corresponding placebo are normally unknown even for the prescribing medical doctor. We are left with the disappointing realization that one cannot say anything about the efficacy of many prescribed drugs.
\begin{figure}[H]
\includegraphics[scale=1.3]{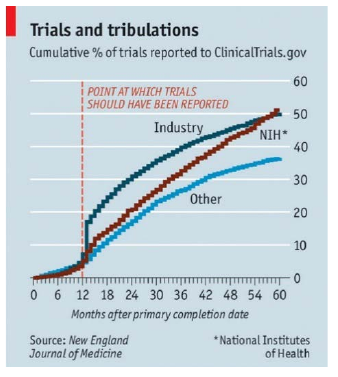}
\caption{Cumulative percentage of test results published after completion, Source: The Economist, 25 July 2015.}
\label{fig:mathematics-1159857-f003}
\end{figure}

This leads us to the following conclusions: 

\begin{enumerate}[label=$\bullet$]
\item A significance of 95\% is required almost everywhere but does by no means represent a proven optimum.
\item Efficacy and statistical significance alone leave a lot of space for interpretation, especially because these numbers are rarely reported, not to mention the complete experimental data.
\item To achieve a certain efficacy with 95\% statistical significance, it may suffice to simply repeat the study about 20 times, i.e., without adjusting the pharmaceutical formulation at all.
\end{enumerate}

\subsection{The False-Positive Problem \label{sect:sec5dot4-mathematics-1159857}}

Eventually, let us address the problem of false-positive results which concerns most empirical sciences and might be one of the main reasons why a considerable number of published results turn out to be false, even if fraud is not prevalent. The following is far from being a new discovery---even YouTube videos~\cite{B20-mathematics-1159857} have been published on this matter. Unfortunately, this common knowledge is not so common at all. We will refer to an example from venture capital in order to illustrate how the resulting implications even affect the finance industry.

Venture capital (VC) describes a type of investment, where money is gathered in order to fund projects which promise an immense return on investment (ROI) but also come with a high failure risk. For simplicity, let us assume a \emph{high return/total loss} scenario and nothing in between. An exaggerated set of numbers has been deliberately chosen to elucidate the point. We assume that 1000 projects apply for investment at a venture capital fund. All these projects promise to return ten times the invested capital within five years which corresponds to an ROI of 900\%. Probably only 1\% of the above mentioned 1000 projects will have a true potential for such a high return (It is also possible to say that 10\% have such a potential. Due to chaos~\cite{B15-mathematics-1159857,B16-mathematics-1159857} it is impossible to predict whether this potential will be realized. The realization rate may also be 10\%). Identifying these ten projects in advance is impossible. The core competency of VC companies is to identify as many of the truly profitable projects as possible. For a simple computational example, we assume that a VC company can invest in ten projects, and that two of these will yield an ROI of 900\% while eight will fail:

\begin{equation}
\label{eq:FD30-mathematics-1159857}
\text{ROI}_{\text{eff}} = \frac{2\text{$\cdotp $}900\% - 8\text{$\cdotp $}100\%}{10} = 100\%.
\tag{27}
\end{equation}

Assuming continuous compounding and an ROI of 100\% within 5 years, we find for the annual interest rate

        \begin{equation}
\label{eq:FD31-mathematics-1159857}
\frac{\ln 2}{5}\text{$\cdotp $}100\%~ \approx 13.9\%.
\tag{28}
\end{equation}

This is a high return even though the VC company in this example identified only two out of ten truly well-performing projects. In other words, such an investment seems to be a good bet, at least at first sight. Let us consider another exemplary constellation. Even if the VC company selects all promising-looking projects out of the 1000 initially given ones with an accuracy of 90\%, it will only find nine of the ten truly performant projects. This alone is not too bad, but in the same run, 99 of the 990 potential failures (10\%) will also be chosen. They are false positives, which will lead to the following effective ROI:\begin{equation}
\label{eq:FD32-mathematics-1159857}
\text{ROI}_{\text{eff}} = \frac{9\text{$\cdotp $}900\% - 99\text{$\cdotp $}100\%}{108} \approx - 16.67\%.
\tag{29}
\end{equation}

\textls[-20]{Hence, within 5 years we end up with an effective annual interest rate of $ln\left( {1 - 0.1667} \right)/$} $5\text{$\cdotp $}100\% \approx - 3.65\% $. In any case, venture capital is far from creating a constant annual return. It will have years of tremendous gains (which most people like to remember) and years of losses (which most people try to forget). A (slight) loss in the long run, and hence the false-positive selection, will be hardly noticed.

The selection of promising projects in empirical sciences comes with the same problem: it may be hard to distinguish truly promising projects from only promising-looking publications. This becomes particularly clear in pharmaceutical research. Identifying, for instance, 10 substances out of 1000 which are potentially worth investigation in clinical trials is the goal of pharmaceutical research. As illustrated in the above discussion, there is a significant risk of identifying more ineffective substances than truly useful ones. In order to counteract this risk, some people suggest that a falsification should also be taken into account and published as a research result. However, finding a project that does not lead to a positive result is easy and would probably lead to an even higher number of publications than already present where one cannot judge about the true value of the research.

\section{Discussion and Conclusions \label{sect:sec6-mathematics-1159857}}

Statistical significance appears to be a problem that has been solved long ago. In more recent publications such as~\cite{B5-mathematics-1159857} only its smooth processing is emphasized and discussed. The prototypical problem is the binominal distribution of, for instance, the efficacy of a medication compared to the (fixed) efficacy of a placebo. If the efficacy of the medication is with a certain significance higher than the placebo efficacy, the medication is approved.

As the efficacy of the placebo is determined from a statistical experiment in the same way as the medication efficacy, it will show a binomial distribution as well. Taking both distributions into account will lead to a two-dimensional probability distribution. As discussed, the determination of the statistical significance requires a double sum or double integral in this case. Compared to the assumption of a constant placebo efficacy, more experiments are needed to obtain the same statistical significance, or a lower significance for a given number of experiments is computed.

The implications, in particular for medication testing, are distressing. Depending on the individual test data, many officially approved medications may in fact not show the required 95\% significance. In \sect{sect:sec3-mathematics-1159857} we have shown that one may need more than twice as many experiments to obtain the required significance. This means that some pharmaceutical research projects may have to be extended, but also that some presently approved medications need further testing.

In the next step, real data from medication tests should be considered as an example. As such data are rarely published and can therefore be hard to find, one may alternatively present tables with more values than in the single examples in this publication. This can be rather time consuming. Note also that it seems impossible to create an Excel tool for performing these calculations, as they need a lot of computational resources.

\vspace{6pt}
\authorcontributions{M.T., G.K., and M.G. contributed to conceptualization, formal analysis, investigation, methodology, writing original draft, writing review and editing. All authors have read and agreed to the published version of the manuscript.}
\funding{This research received no external funding.}
\institutionalreview{Ethical review and approval were waived for this study because it does not involve any experiments with animals or humans.}
\dataavailability{This publication did not use any data not published within the paper or its references.}
\acknowledgments{The authors wish to thank Andr{\fontencoding{T5}\selectfont{\'e}} Grabinski for review and valuable input \mbox{and also~\cite{B12-mathematics-1159857}}.

}
\conflictsofinterest{The authors declare no conflict of interest.}
\end{paracol}

\reftitle{References}

\end{document}